\documentclass{cmip-l}
\usepackage{amssymb,epsf}
\newcommand{\be}{\begin{equation}}
\newcommand{\ee}{\end{equation}}
\newcommand{\bea}{\begin{eqnarray}}
\newcommand{\eea}{\end{eqnarray}}
\newcommand{\beas}{\begin{eqnarray*}}
\newcommand{\eeas}{\end{eqnarray*}}

\def\ft{\;\raisebox{-2mm}{\epsfysize=6mm\epsfbox{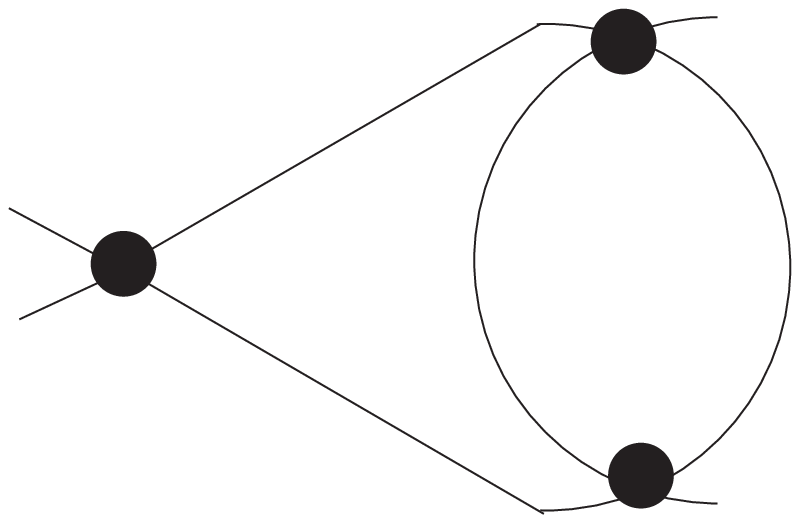}}\;}
\def\tf{\;\raisebox{-2mm}{\epsfysize=6mm\epsfbox{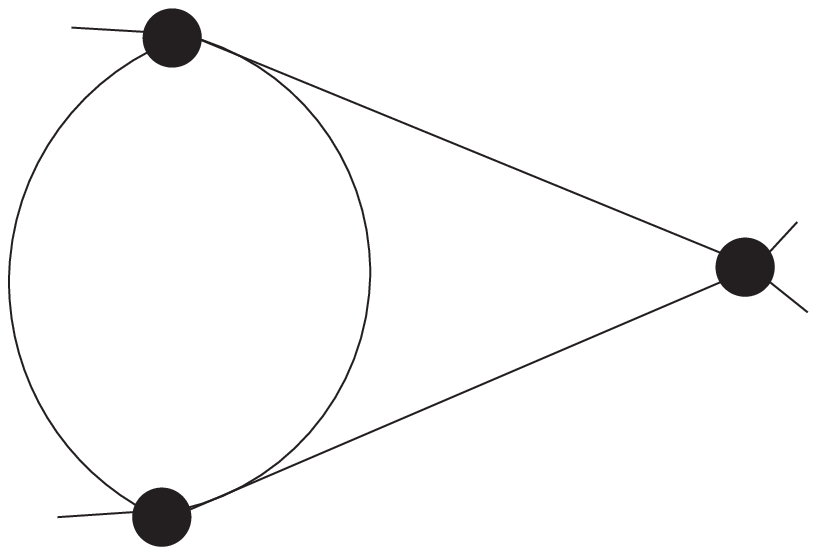}}\;}
\def\fo{\;\raisebox{-2mm}{\epsfysize=6mm\epsfbox{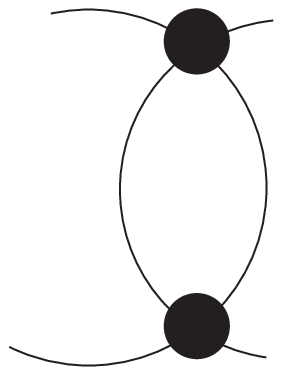}}\;}
\def\foo{\;\raisebox{-2mm}{\epsfysize=6mm\epsfbox{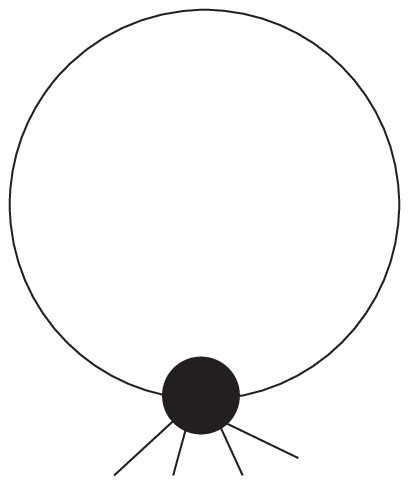}}\;}
\def\so{\;\raisebox{-2mm}{\epsfysize=6mm\epsfbox{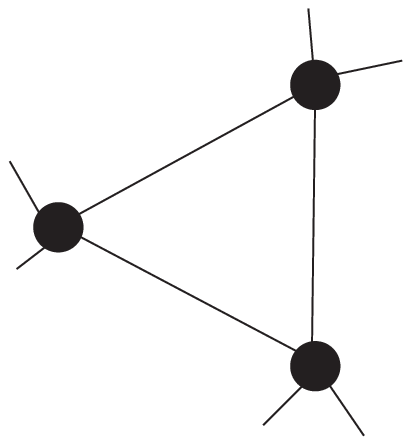}}\;}
\def\wcuts{\;\raisebox{-6mm}{\epsfysize=16mm\epsfbox{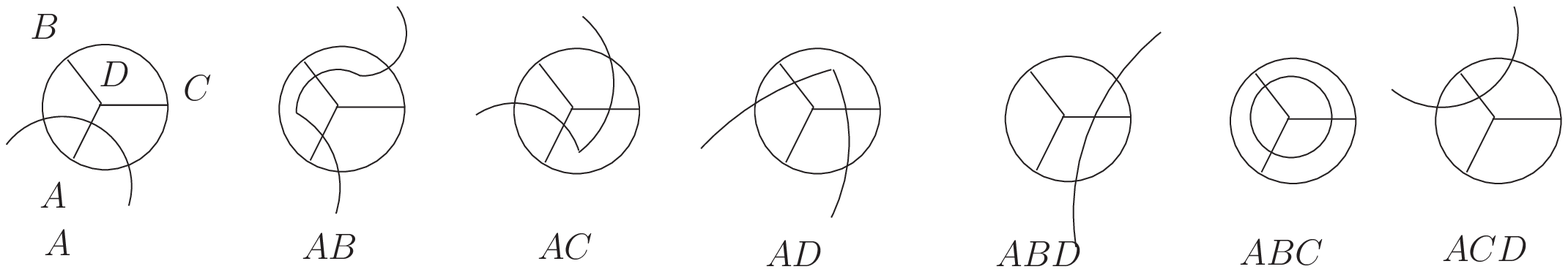}}\;}
\def\wthree{\;\raisebox{-6mm}{\epsfysize=16mm\epsfbox{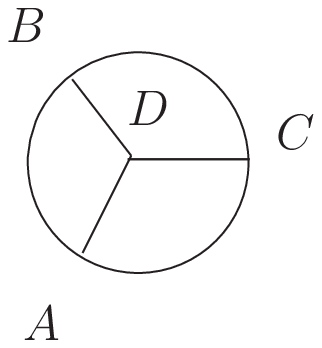}}\;}
\def\wt{\;\raisebox{-2mm}{\epsfysize=6mm\epsfbox{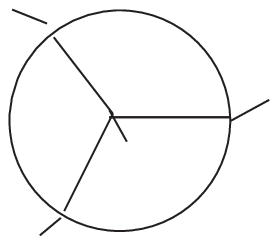}}\;}
\def\st{\;\raisebox{-2mm}{\epsfysize=6mm\epsfbox{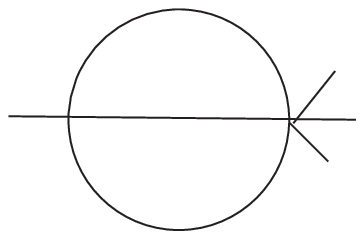}}\;}
\def\et{\;\raisebox{-2mm}{\epsfysize=6mm\epsfbox{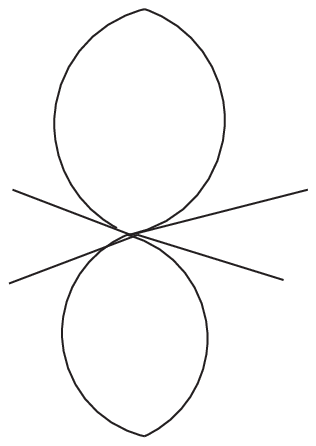}}\;}
\def\eo{\;\raisebox{-2mm}{\epsfysize=6mm\epsfbox{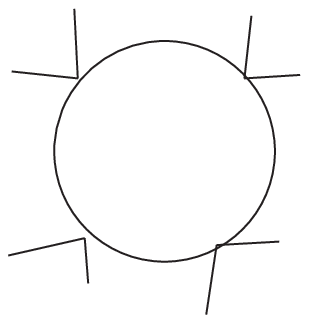}}\;}
\def\sts{\;\raisebox{-2mm}{\epsfysize=6mm\epsfbox{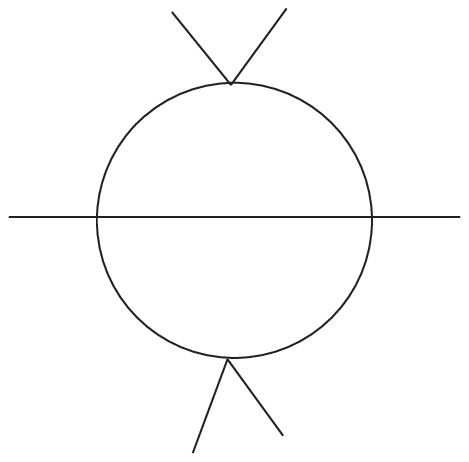}}\;}
\def\tto{\;\raisebox{-2mm}{\epsfysize=6mm\epsfbox{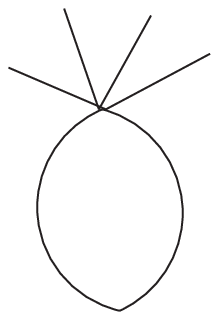}}\;}
\def\dsecore{\;\raisebox{-24mm}{\epsfysize=60mm\epsfbox{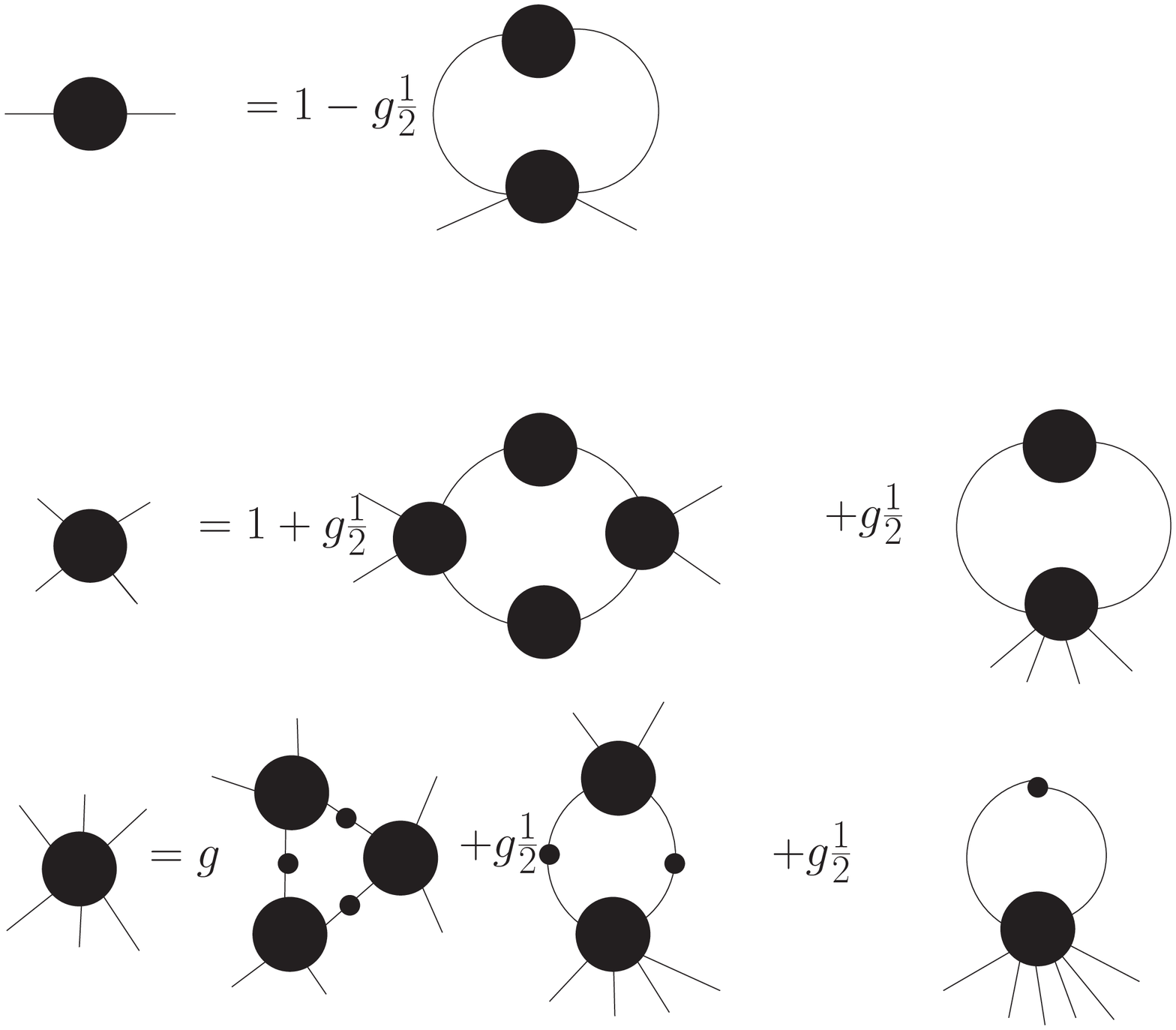}}\;}
\def\saus{\;\raisebox{-2mm}{\epsfysize=6mm\epsfbox{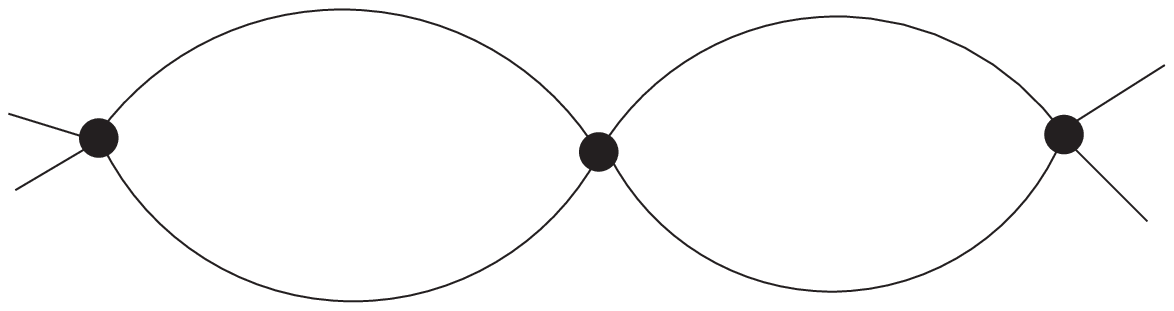}}\;}

\def\One{\mathbb{I}}

\title{The core Hopf algebra}
\author{Dirk Kreimer}
\address{kreimer@ihes.fr, IHES, 35 rte. de Chartres, 91440 Bures-sur-Yvette, France (http://\ www.ihes.fr)
and Boston U.\ (http://math.bu.edu)}
\thanks{${}^*$Contributed to the proceedings for Alain Connes. Work supported in parts by grant NSF-DMS/0603781. Author supported by CNRS}
\begin{document}
\maketitle
\begin{abstract}
We study the core Hopf algebra underlying the renormalization Hopf algebra.
\end{abstract}
\section{Introduction}
In a recent study of the role of limiting mixed Hodge structure, Spencer Bloch and the author introduced the core Hopf algebra
on one-particle irreducible graphs (1PI graphs, also dubbed core graphs in \cite{BlKr}). It is a Hopf algebra which contains the renormalization Hopf algebra as a quotient algebra. One can also view it as the renormalization Hopf algebra of a field theory formulated in infinite dimension, as then any graph which has closed loops is superficially divergent, and any sum over all superficially divergent 1PI graphs reduces to a sum over 1PI graphs.

In this short contribution, we introduce the core Hopf algebra in examples and discuss its larger role in quantum field theory. Formal proofs are to be found in future work. Our main task is to outline some intriguing aspects of the Hopf algebra structure underlying perturbation theory,
going far beyond the problem of renormalization. I feel these ideas are a fitting tribute to my earlier papers with Alain on the subject \cite{CK,RHI,RHII}, and I report on these ideas here for the first time in public in deep respect for Alain's contributions to science, and in deep gratitude for his friendship.
\section{The core Hopf algebra}
The basic formula for the Hopf algebra of a renormalizable field theory is
\be \Delta(\Gamma)=\Gamma\otimes \One +\One \otimes \Gamma+\sum_{\gamma=\cup\gamma_i,\omega(\gamma_i)\leq 0}\gamma\otimes\Gamma/\gamma.\ee
Here, the sum runs over disjoint unions of superficially divergent one-particle irreducible graphs, and $\Gamma/\gamma$
is obtained by shrinking in $\Gamma$ each component $\gamma_i$ of $\gamma=\cup_i\gamma_i$ to a point.
A component $\gamma_i$ is superficially divergent if $\omega(\gamma_i)={\mathrm b}(\gamma_i)D-w(\gamma_i)\leq 0$. Here, ${\mathrm b}$ gives the first Betti number, the number of independent cycles,
$D$ is the dimension of spacetime and $w(\gamma_i)$ the sum of the scaling weights of internal edges and vertices of $\gamma_i$.
See \cite{BlKr} for notation and details. This renormalization Hopf algebra can be easily augmented to take care of the quantum numbers which label external legs, incorporating formfactors and kinematics of Feynman amplitudes. We focus here on some elementary aspects of iteration of subgraphs into each other, and will not clutter notation any further.

The core Hopf algebra is then obtained by relaxing the qualification on superficial divergence: we simply sum over all 1PI subgraphs.
\be \Delta(\Gamma)=\Gamma\otimes \One +\One \otimes \Gamma+\sum_{\gamma=\cup\gamma_i}\gamma\otimes\Gamma/\gamma.\ee
Note that this immediately implies that the only primitives are one-loop graphs. As an aside, we note that for the renormalization Hopf algebra of quantum gravity, the particular powercounting rules of gravity \cite{grav} ensure that for perturbative gravity, the 
renormalization Hopf algebra and the core algebra agree.

Let us give now an example for the core Hopf algebra in $\phi^4$ theory.
\bea \Delta_c\left(\ft\right) & = & \ft\otimes\One+\One\otimes\ft\nonumber\\
 & & +2\so\otimes\foo+\fo\otimes\fo.\eea
In the renormalization Hopf algebra we would simply have
\be \Delta\left(\ft\right)  =  \ft\otimes\One+\One\otimes\ft+\fo\otimes\fo.\ee
So why shall we study the core Hopf algebra?  Let us discuss the structure of the graph polynomial:
\be \phi(\Gamma)=\sum_{{\rm spanning\; trees}\;T}\prod_{e\notin T}A_e,\ee
 accompanying this graph.
Labeling the two straight edges on the left as $A_1,A_2$ and the other two as $A_3,A_4$, it reads 
\bea \phi\left(\ft\right) & = & A_1A_3+A_1A_4+A_2A_3+A_2A_4+A_3A_4\\
 & = & (A_1+A_2)(A_3+A_4)+A_3A_4,\label{fac1}\\
 & = & (A_1+A_2+A_3)A_4+(A_1+A_2)A_3\label{fac2}\\
 & = & (A_1+A_2+A_4)A_3+(A_1+A_2)A_4\label{fac3}\eea
corresponding to the five spanning trees of the graph.
We can find the coproduct of the renormalization as well as the core Hopf algebra from a factorization
\be \phi(\Gamma)=\phi(\Gamma/\gamma)\phi(\gamma)+r(\Gamma,\gamma)\ee
such that $r(\Gamma,\gamma)$ is of higher degree in the variables of $\phi(\gamma)$ than $\phi(\gamma)$ itself.
For example, from (\ref{fac1}) 
\be \phi\left(\ft\right)=\phi\left(\fo\right)\phi\left(\fo\right)+A_3A_4,\ee
where $A_3A_4$ is quadratic in the variables $A_3,A_4$ of the subgraph made of edges $3,4$ of the initial graph,
while that subgraph $\gamma$ itself, superficially divergent as $\omega(\gamma)=0$,  has graphpolynomial
\be \phi\left(\fo\right)=A_3+A_4,\ee
while the cograph has
\be \phi\left(\ft/\fo\right)=A_1+A_2.\ee
Clearly, when $A_3,A_4$ tend to zero jointly, $r(\Gamma,\gamma)$ vanishes faster than $\phi(\Gamma/\gamma)\phi(\gamma)$
and hence we find a subdivergence with regard to the $A_3,A_4$ integration using the Feynman rules in parametric representation.
The other factorizations (\ref{fac2},\ref{fac3}) above have limits which remain integrable over the respective subgraph variables.
 
But any investigation of the algebra-geometric structure of periods assigned to graphs starts 
with the investigation of the graph hypersurface $X_\Gamma: \phi(\Gamma)=0$, and the question how that graph hypersurface meets the simplex $A_i>0$. Integrability is a rather irrelevant criterion in this respect,
as studied in detail in \cite{BlKr}, and the two other factorizations (\ref{fac2},\ref{fac3})
\beas A_1A_3+A_1A_4+A_2A_3+A_2A_4+A_3A_4 & = & \underbrace{(A_1+A_2+A_3)A_4}_{\rm{linear\; in} A_4}+\underbrace{(A_1+A_2)A_3}_{{\rm constant\; in} A_4}\\ & = & \underbrace{(A_1+A_2+A_4)A_3}_{\rm{linear\; in} A_3}+\underbrace{(A_1+A_2)A_4}_{\rm{constant\; in} A_3},\eeas
give the other two terms generated by the non-trivial part of $\Delta_c$
and are mandatory to study the situation from the perspective of a limiting mixed Hodge structure.
Note that \be \omega\left(\so\right)=+1.\ee

As an amusing side remark, let me mention that the famous problem of overlapping divergences in renormalization
corresponds to precisely the coexistence of different factorizations in the above sense. While the above has three coexistent decompositions of the graph polynomial all contributing to the core coproduct, only a single term contributes to the renormalization coproduct as this graph has no overlapping divergences with regard to renormalization.

For the renormalization Hopf algebra it has proved worthwhile to study its Hochschild cohomology, as this provides a prefered way to prove renormalizability of counterterms and illuminates the structure of Dyson Schwinger equations (DSE). Let us see how the core Hopf algebra fares in this respect.
\section{DSE in the core Hopf algebra}
Let us stay for simplicity in the realm of massless $\phi^4$ theory in four dimensions of space time.
In the renormalization Hopf algebra, we have to study  two Green functions, one for the vertex function (four external legs), and one for the inverse propagator (two external legs).

Both are obtained from the evaluation by suitably renormalized Feynman rules $\phi_R$ of the series $X^4(g)$ and $X^2(g)$ of all 1PI graphs with the appropriate number of four or two external legs.

These series in the coupling $g$ (series in $g$ with coefficients in the Hopf algebra)
are fixpoints of equations formed by studying the Hochschild cohomology \cite{BergbKr}, $bB_+^{j,m}=0$, $m\in\{2,4\}$ of these Hopf algebras:
\bea X^4(g) & = & \One+\sum_{j>0} g^j B_+^{j,4}\left(X^4(g)\left(\frac{X^4(g)}{(X^2(g))^2}\right)^j\right)\label{DSE4}\\
X^2(g) & = & \One-\sum_{j>0} g^j B_+^{j,2}\left(X^2(g)\left(\frac{X^4(g)}{(X^2(g))^2}\right)^j\right).
\eea
It is crucial that the $B_+^{j,m}$ are closed one-cocycles: it leads to a clean approach to non-perturbative aspects of local field theory and to an analysis of the structure of solution of Dyson--Schwinger equations in such theories \cite{KY1,KY2,KY3}.

In the above, \be B_+^{j,m}=\sum_{|\gamma|=j,\Delta(\gamma)=\gamma\otimes\One+\One\otimes\gamma}\frac{1}{{\rm sym}(\gamma)}B_+^{\gamma},\ee
with $\gamma$ having $m$ external legs and
\be
B_+^\gamma(X)=\sum_{\Gamma}\frac{{\mathrm{bij}}(\gamma,X,\Gamma)}{\mathrm{maxf}(\Gamma)[\gamma|X]|X|_\wedge}\Gamma.\label{bpdef}
\ee
The reader will have to consult \cite{anatomy,KYthesis} for details.
We just mention that ${\rm bij}(\gamma,X,\Gamma)$ counts the number of bijections between external edges of $X$ and insertion places of $\gamma$ so as to obtain $\Gamma$, $\rm maxf$ counts the number of ways to shrink 1PI subgraphs such that the cograph is primitive under the coproduct, $[\gamma|X]$ counts the number of insertion places for $X$ in $\gamma$, and $|X|_\wedge$ gives the number of different graphs generated from permuting external edges.

Here is an illuminating example: 
First, from Hochschild closedness, $B_+^\gamma(\One)=\gamma,\forall \gamma$.
Hence $X^4(g)$ starts as 
\be \One+\frac{1}{2}\fo+\ldots+{\mathcal O}(g^2).\ee
Here, $+\ldots$ refers to the two other orientations of this graph (the $s,t,u$ channels).
Let us now look at 
\be \frac{1}{2}B_+^{\fo+\ldots}\left([X^4(g)]^2/[X^2(g)]^2\right)\ee
appearing on the rhs of (\ref{DSE4}). 

Let us Taylor expand the argument in $g$ to first order and concentrate at the term coming from the expansion of the square of the vertex function. 
We find
\be \frac{1}{2}B_+^{\fo+\ldots}\left(2\times \frac{1}{2}\times \fo+\ldots\right).\ee
The number of insertion places is 
\be \left[\fo|\fo\right]=2,\ee
there are three orientations,
\be \left|\fo\right|_\wedge=3,\ee 
giving two bijections leading to graphs of the form 
\be \ft,\ee (swapped or permuted possibly) each of which has a single maximal forest, maxf$=1$,
and one bijection leading to 
\be \saus.\ee with two maximal forests.

We hence find
\beas \frac{1}{2}B_+^{\fo+\ldots}\left( \fo+\ldots\right) & = & \frac{1}{4} \left[  \saus+\ldots\right]\\ 
& & +\frac{1}{2}\left[\ft+\tf\ldots\right],\eeas
with all the correct symmetry factors.
Computing now the coproduct delivers 
\beas
\Delta\left(\frac{1}{2}B_+^{\fo+\ldots}\left(\fo+\ldots\right)\right) & = & 
\frac{1}{2}B_+^{\fo+\ldots}\left(\fo+\ldots\right)\otimes \One\\
 & & +\One\otimes \frac{1}{2}B_+^{\fo+\ldots}\left(\fo+\ldots\right)\\
 & & + \frac{1}{2}\left[\fo+\ldots\right]\otimes [\fo+\ldots]
\eeas
which agrees with 
\beas 
\frac{1}{2}B_+^{\fo+\ldots}\left( \fo+\ldots\right)  \otimes  & &  \One\\
+\left({\rm id}\otimes \frac{1}{2}B_+^{\fo+\ldots} \right)\Delta\left(\left[\fo+\ldots\right]\right),
\eeas
as required by Hochschild cohomology. Hochschild cohomology does us an enormous favor here, and it becomes even more impressive when one realizes how it conspires to give rhyme and reason to internal symmetries in a field theory \cite{anatomy,w1,w2}.

So what changes if we try the same with the core Hopf algebra?
 
Let us describe first the primitives. We noted already they are all one-loop graphs.
Next, we observe that in the core Hopf algebra underlying the vertices and edges of $\phi^4_4$ theory we must have vertices of
arbitrary high but even valence. 

Hence, the one-loop graphs can be described by partitions, where each entry
$j$ in the partition corresponds to a $2j+2$ valent vertex in the one-loop graph, the length of the partition gives the number of vertices on the one-loop graph (and equals the number of its edges), and the size of the partition gives the total number of external edges.

So for example 
\be \fo\sim (1,1)\ee
and 
\be \so\sim (1,1,1).\ee

We are led to the following system:
\bea X^2(g) & = & \One-gB_+^{(1)}(X^4/X^2(g)),\\
X^4(g) & = & \One+g\frac{1}{2}B_+^{(1,1)}([X^4]^2/[X^2(g)]^2)+g\frac{1}{2}B_+^{(2)}(X^6(g)/X^2(g)),\\
X^6(g) & = & gB_+^{(1,1,1)}([X^4(g)]^3/[X^2(g)]^3)\\ & & +g\frac{1}{2}B_+^{(2,1)}([X^6(g)X^4(g)]/[X^2(g)]^2)\nonumber\\ & & +g\frac{1}{2}B_+^{(3)}(X^8(g)/X^2(g)),\nonumber
\eea
and so on,
which is best understood graphically (we omit to give contributions obtained by swapping or permuting external edges):
$$\dsecore.$$
Note that we have only a finite number of one-loop primitives contributing to each fix-point equation, but we have an infinite set of equations to consider. Also, we emphasize that we maintain the $B_+$ operators to be closed one-cocycles in the Hochschild cohomology
of the core Hopf algebra, and claim that the same definition (\ref{bpdef}) achieves precisely that.

The series $X^4$ and $X^2$ which are fixpoints of the above system are the same series as the one obtained in the Hochschild cohomology
of the renormalization Hopf algebra above. This is a rather remarkable fact. We have done something very typical for the functional integral actually: we have traded a loop expansion for a leg expansion.

It is instructive to see in an example how this comes about. From $B_+{(1,1)}$ we get the same graphs as before, but 
\be \ft\ee has now three maximal forests (in the core Hopf algebra the number of maximal forests equals the number of non-self-intersecting closed paths we can draw on the graph).
So this contribution gets an extra factor $1/3$. The missing $2/3$ is precisely provided from the same graph generated by insertions of
\be \so\ee into $B_+^{(2)}$, where the number of relevant bijections is two.

\section{sub Hopf algebras and AdS/CFT}
Now, for the renormalization Hopf algebra and its Hochschild cohomology we have learned a rather remarkable story:
if we decompose a series of graphs by order, 
\be X^s(g)=\sum_{j=0}^\infty c_j^s g^j,\ee
with $c_j^s$ Hopf algebra elements, these finite linear combinations of graphs provide a sub Hopf algebra.
To achieve this in the presence of internal symmetries one has to divide by  suitable ideals \cite{anatomy,w1,w2}, and doing so,
we finally can work with much simpler Hopf algebras. Combining with the structure of the renormalization group \cite{RHII,KY1,KY2,KY3} then fully exhibits the recursive structure of field theory parameterized by the periods underlying the motives coming with the graph hypersurfaces.

And for the core Hopf algebra?  If we sum all graphs contributing to a chosen amplitude at a given loop order, form these linear combinations the generators of a sub Hopf algebra? Certainly not as they stand, but what is the structure of the (co-) ideals such that we can obtain such a sub Hopf algebra when taking quotients?

Applying the techniques of \cite{anatomy,w1,w2} this is straightforward as we will report elsewhere \cite{KvS}.
The harder question is to study Feynman rules and see to what extent they respect such quotients.

Here, we note that the relations \be X^{2k}/X^{2(k-1)}=X^{2(k+1)}/X^{2k}\label{tree}\ee determine a co-ideal 
such that we get the desired sub Hopf algebras. Similar relations will show up in the study of any core Hopf algebra for other quantum field theories.

Two points deserve attention: if we had not considered $\phi^4$ but perturbative gravity, these would be precisely the relations which,
if tolerated by the Feynman rules, will render gravity renormalizable.  

Furthermore, at tree level, the relations (\ref{tree}) have a recursive form very familiar from studying the now famous \cite{onshellrec}
on-shell recursion relations of tree (and actually one-loop) amplitudes. This deserves much more attention in the future.
Note in particular that one-loop recursion relations boil down here to
relations between the Hochschild one-cocycles driving the equations of motion.
\section{Unitarity of the $S$-matrix}
The main role which the core Hopf algebra has to play in the future is, I believe,
in reconciling our understanding of renormalization with the unitarity of the $S$-matrix.
The notion of a cut at a Feynman graph is compatible with the core coproduct. This again will be discussed elsewhere,
but let us give us one example.
Consider the wheel with three spokes
\be \wthree.\ee
We have labelled its vertices $A,B,C,D$. External edges are not drawn, but all vertices are supposed to be four-valent.
We consider the graph as contributing to a $1\to 3$ production amplitude, and consider the particle incoming at vertex $A$.
The core Hopf algebra delivers the following coproduct for this graph:
\beas \Delta_c\left(\wt\right) & = & \wt\otimes\One+\One\otimes\wt\\
 & & +4\so\otimes \st+3\eo\otimes \et+\sts\otimes\tto.\eeas
Note that from the terms on the rhs only $\st$ allows for cuts $C$ separating incoming and outgoing particles.

The other ones are too tadpole-ish to contribute:
\be C\left(\et\right)=C\left(\tto\right)=0.\ee
Now  consider the cuts $C$ determining the imaginary part.
\be C\left(\wt\right)=\wcuts.\ee
We see four contributions which have an intact subgraph $\so$, and three contributions where no internal loop is left intact.
We have labeled each cut by the set of vertices connected to vertex $A$.

If we now let $CC$ (completely cut) be the operator which assigns the sum of all cuts to a graph such that no internal loop is left intact, then 
\be ({\rm id}\otimes CC)\Delta_c\left(\wt\right)\ee is in one-to-one correpondence with $C(\wt)$ and hence
describes the structure of this imaginary part rather well.

This is the beginning of a mathematically beautifully approach to unitarity
and the $S$-matrix based on the core Hopf algebra. I hope to report more on that in collaboration with Spencer Bloch,
celebrating still a line of thought which started in \cite{K} and first blossomed in my work with Alain.

\end{document}